\documentclass[journal]{IEEEtran}

\IEEEoverridecommandlockouts

\usepackage{cite}

\usepackage{amsmath,amssymb,amsfonts}

\usepackage{algorithm}

\usepackage{algorithmicx}

\usepackage{graphicx}

\usepackage{textcomp}

\usepackage{xcolor}

\usepackage[justification=centering]{caption}

\usepackage{algpseudocode}

\makeatletter

\newcommand{\removelatexerror}{\let\@latex@error\@gobble}

\makeatother

\usepackage{comment}

\usepackage{booktabs}

\usepackage{soul}

\soulregister\cite7

\soulregister\citep7

\soulregister\citet7

\soulregister\ref7

\soulregister\pageref7

\def\BibTeX{{\rm B\kern-.05em{\sc i\kern-.025em b}\kern-.08em

    T\kern-.1667em\lower.7ex\hbox{E}\kern-.125emX}}

\begin{document}

\title{Precise WiFi Indoor Positioning using Deep Learning Algorithms}

\author{ Minxue Cai and Zihuai Lin

School of Electrical and Information Engineering, The University of Sydney, Australia

Emails:  zihuai.lin@sydney.edu.au.


}

\maketitle
\bibliographystyle{ieeetr} 

\begin{abstract}

This study demonstrates a WiFi indoor positioning system using Deep Learning algorithms.  A new method using fitting function in MATLAB will be utilized to compute the path loss coefficient and log-normal fading variance. To reduce the error, a new hybrid localization approach utilizing Received Signal Strength Indicator (RSSI) and Angle of Arrival (AoA) has been created. Three Deep Learning algorithms would be utilized to decrease the adverse influence of the noise and interference. This paper compares the performance of two models in three different indoor environments. The average error of our hybrid positioning model trained by CNN in the big classroom is less than 250 mm. 

\end{abstract}


\maketitle

\section{Introduction}

\label{sec:introduction}

The Internet of Things (IoT) and mobile networks are changing people’s lives and are also contributing to the development of indoor positioning and indoor navigation technologies \cite{leng2022design}.

Compared with indoor positioning, outdoor positioning technology is now more mature. Global Positioning System (GPS) requires open skies to receive satellite signals and it needs to receive at least four satellite signals for accurate positioning. Therefore, a large number of indoor obstructions can prevent GPS systems from receiving sufficient satellite signals, resulting in significant positioning errors or the inability to locate accurately. In addition, multipath effect and signal fading can also contribute to an increased positioning error in GPS technology \cite{yeh2009study}.

In order to solve the problem of GPS inaccuracy in indoor environments, indoor positioning technology has been developed rapidly and applied in many fields. Among the technologies, WiFi indoor positioning technology is widely utilized. WiFi is a wireless network technology by using the IEEE 802.11 standard. WiFi signals are generally divided into two frequency bands, 2.4 GHz and 5 GHz. Compared with the signals of 5GHz band, the signals of 2.4GHz band have better penetration ability and can transmit longer distances. Therefore, the signals in the 2.4 GHz band are more suitable for indoor positioning. In addition, devices such as WiFi routers, WiFi access points, and wireless cards can generate WiFi signals, thereby increasing the flexibility of applying WiFi in indoor positioning field \cite{cho2012wifi}. 

To further reduce the negative influence of NLoS problem and multipath effect, Deep Learning algorithms will be utilized in indoor positioning models. The Deep Learning algorithms can help to identify the relationship between the WiFi signal information and the target points. The Deep Learning algorithms can also decrease the negative influence of noise and interference by changing the weights of the neurons in the layers \cite{liu2022human,liu2021machine}. 


The remainder of the paper is organized as follows. Section II reviews the concepts of indoor positioning field and the existing neural network-based indoor localization methods. Section III demonstrates the system model and positioning methods used in the paper. Section IV describes the system model and the methods used in this paper.  Section V analyzes the experimental results. The last Section gives a summary of this work and the
future work plan.
\section{Background}\label{sec:background}
\subsection{WiFi Ranging-based Technology}
According to different technical principles, WiFi indoor positioning technology can be divided into Angle of Arrival (AoA), Received Signal Strength Indicator (RSSI), and Time of Flight (ToF) \cite{zhai2015rss}.

The Received Signal Strength Indicator is a method by utilizing the signal strength at the receiver point. The energy loss during propagation will be calculated by comparing the signal strength of the receiving and transmitting points. It will be used as an input to the log-distance path loss model to compute the distance from the transmitting point to receiving point. As a result, the location of the receiving point in the indoor environment will be obtained \cite{achroufene2018rss}.

The Angle of Arrival method is a technique that utilizes the angle of arrival of the signal to calculate the target position. By deploying a large number of antenna arrays in an indoor environment, phase differences between different antennas will be used to calculate the angle of arrival of the signal \cite{zhong2014cooperative}.

Time of Flight is a technique that uses the propagation time of a wireless signal to calculate the location of a target. The transmitter sends a signal to the receiver, and the receiver receives the signal and immediately transmits back a response signal. The distance from the sender to receiver is calculated by recording the round-trip time of the signal.The location of the receiver is calculated based on these distances \cite{banin2013next}.

\subsection{Trilateration Method}
RSSI, AoA and ToF methods usually need at least three anchor points to complete the measurement of the target point location \cite{wang2022precise}. Therefore, the Trilateration method is utilized in the field of indoor positioning for ranging-based methods. There are three non-collinear signal transmitters and one unknown receiver in the plane. The distances from the three transmitting points to the receiving point can be calculated. When the coordinates of the three transmitting points are known, three circles can be drawn with the transmitting point as the center and the calculated distance as the radius. Theoretically, the intersection point of the three circles represents the receiving point. However, the positioning accuracy decreases due to signal fading and environmental interference in real indoor environment. Therefore, there is not necessarily only one intersection point of the three circles in the real environment \cite{mahiddin2012indoor}.

\subsection{Deep Learning and Neural Network}
Deep Learning algorithms are developed from Machine Learning algorithms. Deep Learning algorithms are used to study and find out features from data to perform classification, regression or other prediction tasks so that patterns in the data can be identified and classified. A neural network is the core component of Deep Learning. It is a hierarchical structure consisting of multiple neurons, each receiving input from a neuron in the previous layer and passing the output to the next layer of neurons. Neurons have many
weights utilized to adjust the transmission intensity of signals from one neuron to another. These weights are adjusted by training the neural network to minimize the prediction error \cite{hu2019machine} .

Back-Propagation Neural Network (BPNN) is a common multi-layer feed forward neural network. It trains the network through Back-Propagation algorithms to find the mapping relationship between input data and corresponding output targets. BPNN has three types of layers including input layers, hidden layers and output layers. Each layer has many neurons, and each neuron has a weight \cite{aljanad2021neural}.

Radial Basis Function (RBF) is commonly used to solve classification and regression problems. RBF utilizes Radial Basis functions to find the relationship between input data and output targets. The linear output layer will be utilized to compute the results \cite{hamedi2019artificial}. It has a similar structure compared with BPNN .

Convolutional Neural Network (CNN) is a Deep Learning algorithm that is skilled in processing multidimensional data. The most important sections of a CNN are the convolutional layers and the pooling layers. In addition, it also includes fully connected layers, activation function, batch normalization, and other sections. CNN has the abilities to automatically find out and learn the characteristics of entered data through convolutional kernels in the convolutional layer. Afterwards, the Back-Propagation algorithm and Random Gradient Descent method will be used for CNN training \cite{tabian2019convolutional}.

\subsection{Indoor Positioning based on Deep Learning}

In \cite{xue2020wifi}, Xue et al. introduced a highly adaptive indoor localization (HAIL) approach which could take advantage of both relative RSSI values and absolute RSSI values to rise the accuracy. BPNN was used in the designed method to calculate the degree of matching between the absolute RSSI values in the database and the measured absolute RSSI values. The results illustrated that HAIL attained a mean absolute error of 0.87 m.

In \cite{meng2018indoor}, Meng et al. focused on RSSI-based WiFi indoor localization. First of all, the collected RSSI values would be pre-processed by utilizing the weighted median Gaussian filtering method to enhance the reliability of the data. The pre-processed data would be utilized to found a database. An improved fast clustering algorithm was utilized to adjust the amount of neurons on the RBF hidden layer and the kernel function parameters of neurons. The Levenberg Marquard algorithm optimized the input RSSI values. The improved RBF algorithm would improve the matching degree between the measured RSSI values and the RSSI values in the database. The mean absolute error of the improved RBF algorithm was 1.421 m, while that of the RBF algorithm was 1.925 m.

In \cite{zhang2019wireless}, Zhang et al. concentrated on the WiFi fingerprint-based method based RSSI. An algorithm combined with the CNN and Gaussian Process Regression (GPR) was introduced. CNN could automatically find out the features of the input RSSI values and decrease the negative effects of noise and interference in the indoor environments. Meanwhile, GRU could decrease the overfitting problem of CNN. The outcome demonstrated that the mean absolute error of CNN was 1.442 m, while that of hybrid algorithm was 1.06 m.
\section{Methods and System Model}\label{sec:methods and system model}

The system model will consist of three fixed anchor points and a moving target point. Fig.1 demonstrates the detailed layout of the system model.

\begin{figure}[H] 
\centering 
\includegraphics[width=0.3\textwidth]{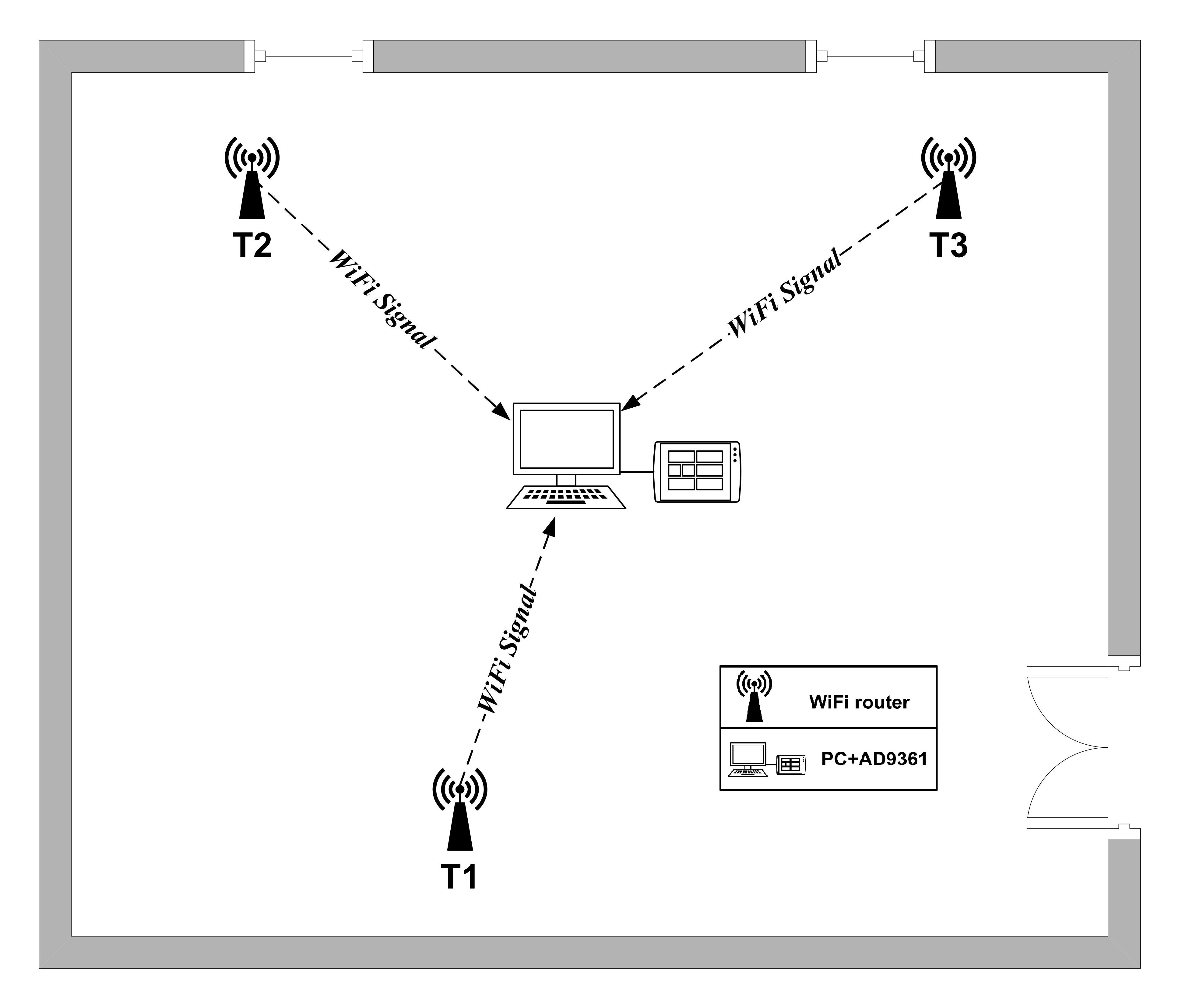}
\caption{The Structure of the System Model} 
\label{Figure_2.1} 
\end{figure}

Three WiFi routers named T1, T2, and T3 are placed as signal transmitters at the fixed anchor points. A laptop and an AD9361 device will be placed as a signal receiver at the target point. The target point will be chosen within the range covered by the three anchor points. All four devices use wireless communication technology based on WiFi signals. Additional functions in MATLAB will be used with the AD9361 instrument to measure the experimental data. To better assess the performance of our designed method, the test environments will be selected from a big classroom, a corridor and a small classroom.

\subsection{RSSI-based Trilateration Method}

The RSSI-based Trilateration method is a common wireless positioning method that calculates the location of a device by measuring the RSSI values. When a WiFi router sends out a signal and a laptop as a receiver receives it, the RSSI value is measured. The position of the receiver is calculated by utilizing the RSSI values of the three transmitters. The log-distance path loss model, which is shown in equation (1),  is utilized to transfer the RSSI value to a distance from the transmitting point to the receiving point.
\begin{equation}
PL(d)=P_{T}(d)-P_{R}(d)=PL(d_{0})+10\gamma\log_{10}\frac{d}{d_{0}}+X_{g} \label{3.1}
\end{equation}
where $PL(d)$ is the total path loss in dBm, $P_{T}(d)$ represents the transmitted power in dBm, $P_{R}(d)$ is the received power in dBm, $d_{0}$ is the reference distance (usually 1 m to 10 m for indoor environments), $PL(d_{0})$ illustrates the path loss in dBm  at the reference distance $d_{0}$, $d$ represents the total length of propagation path, $\gamma$ represents the path loss coefficient and $X_{g}$ represents a Gaussian random variable with zero mean, which means the attenuation caused by fading. Because of the complexity of the indoor scene, NLoS problem and multipath effect have influence on the RSSI values measured at the receiver. Therefore, the random variable $X_{g}$ obeys a Gaussian distribution and has a standard deviation $\sigma$ ,which is also called log-normal fading variance, in decibels \cite{zhang2022machine}. The path loss coefficient and log-normal fading variance vary in different indoor environments. Therefore, a MATLAB fitting function method will be invented to calculate these two parameters.

The equation (1) will become equation (2), which is shown below.
\begin{equation}
P_{R}(d)=P_{R}(d_{0})-10\gamma\log_{10}\frac{d}{d_{0}}-X_{g} \label{3.2}
\end{equation}
In equation (2), $P_{R}(d_{0})$ represents the received power in dBm at the reference distance $d_{0}$. Therefore, the signal strength values at different distances to the signal transmitter need to be measured. In the test environment, the location of the WiFi router is defined as the original point and the reference distance $d_{0}$ will be defined as 1 m.

The functional model of the relationship between the value of RSSI and distance is a sum of a $\log_{10}$ function and a Gaussian distribution function with zero mean. The fitting function in MATLAB will be utilized to fit this functional model. 

The calculated path loss coefficient and the log-normal fading variance will be used in equation (2). The measured RSSI value will be carried into equation (2) to compute the distance from the WiFi router to the receiver point. 

In order to measure the location of the signal reception point more precisely, three WiFi routers will be arranged in each indoor environment. The distances from the three WiFi routers to the receiver will be calculated as $d_{1}$,$d_{2}$ and $d_{3}$ respectively. The three WiFi routers have fixed locations which can be demonstrated as $(x_{1},y_{1})$, $(x_{2},y_{2})$ and $(x_{3},y_{3})$ respectively. 
Three circles are made with three WiFi routers as the centers as well as $d_{1}$,$d_{2}$ and $d_{3}$ as the radii. Their mathematical expressions will be represented by the below expressions.
\begin{equation}
(x-x_{1})^2+(y-y_{1})^2=d_{1}^2 \label{3.3}
\end{equation}
\begin{equation}
(x-x_{2})^2+(y-y_{2})^2=d_{2}^2 \label{3.4}
\end{equation}
\begin{equation}
(x-x_{3})^2+(y-y_{3})^2=d_{3}^2 \label{3.5}
\end{equation}
The least square method will be utilized to compute the coordinates of the receiver point. Equation (3) and equation (4) will subtract equation (5) to obtain the linearized equation
\begin{equation}
AX=b 
\end{equation}
where 
\begin{figure}[htbp] 
\centering 
\includegraphics[width=0.3\textwidth]{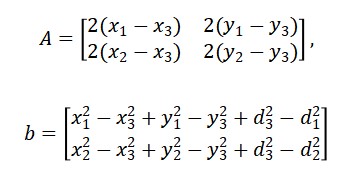} 
\label{Figure_3.3} 
\end{figure}

The coordinates of the receiver point would be computed by utilizing least square approach, which can be illustrated in the below formula.
\begin{equation}
X=(A^TA)^{-1} A^Tb
\end{equation}

There are three WiFi routers in the indoor environment, the multipath effect will prevent the three circles from intersecting at one point. Apart from that, there are some obstacles such as tables and cupboards in the indoor environments. The NLoS problem results in more energy loss during signal propagation, which affects the accuracy of RSSI measurements and indoor localization. Therefore, the coordinates of the calculated receiver point will be located in the common area of the three circles, as presented in Fig.2.

\begin{figure}[H] 
\centering 
\includegraphics[width=0.3\textwidth]{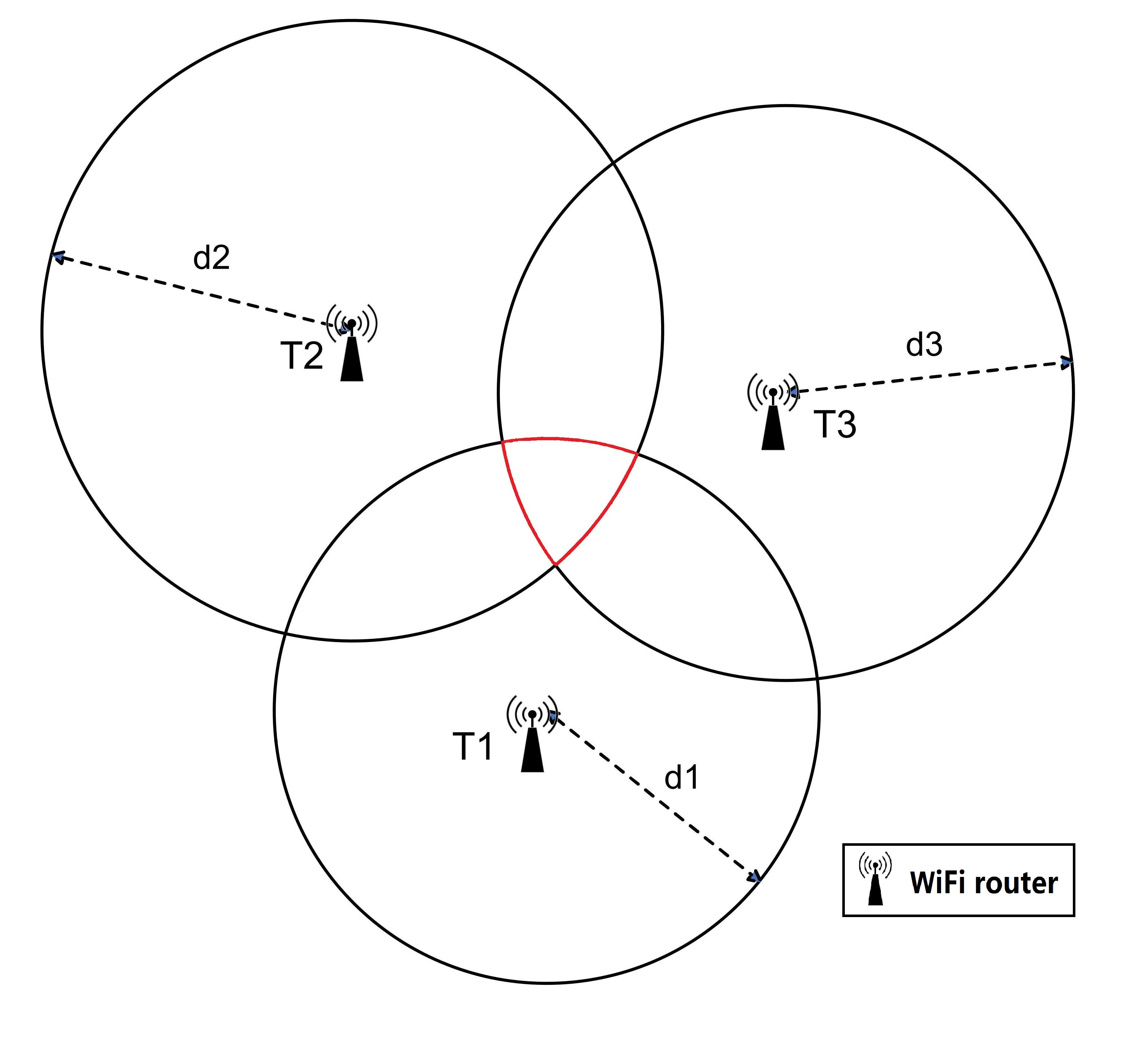} 
\caption{The Result of the RSSI-based Trilateration Method}
\label{Figure_3.3} 
\end{figure}

\subsection{AoA Estimation}
The Multiple Signal Classification (MUSIC) algorithm is a high-resolution spectral analysis method for estimating the direction and number of multiple sources in a signal contaminated with noise \cite{gupta2015music}.The MUSIC algorithm estimates the direction of a signal source primarily through spatial spectral analysis. 

Suppose there is a uniform linear array with $M$ receiving antennas that can receive signals from $K$ sources. For each signal source, its position is assumed to be at $\theta_{k}$, its frequency is $\omega_{k}$, and its sampling time is $T$. The uniform linear array will be used to receive the signal from the source and convert it into a digital signal for processing. So the received signal will have the following expression.
\begin{equation}
x_{i}(t)=\sum_{k=1}^K \alpha_{k}(\theta_{k})e^{j\omega_{k}t}+n_{i}(t), i=1,2,....,M
\label{3.6}
\end{equation}
where $\alpha_{k}(\theta_{k})$ is the spatial filtering coefficient of the signal source $k$ in the direction of $\theta_{k}$ and and $n_{i}(t)$ represents the noise term. The received signal at each receiver can be illustrated by equation (9).
\begin{equation}
x(t)=[x_{1}(t),x_{2}(t),....,x_{M}(t)]^T
\label{3.7}
\end{equation}
Then the received signal vector will be used to calculate the spatial correlation matrix $R$, which can be shown in equation (10).
\begin{equation}
R=\mathbb{E}[x(t)x^H(t)]
\label{3.8}
\end{equation}
$^H$ means the Hermitian transpose and $\mathbb{E}$ represents the expectation operator. Eigenvalue decomposition is performed on the spatial correlation matrix to attain the eigenvalues and eigenvectors:
\begin{equation}
R=U_{R}\Lambda_{R}U_{R}^H
\label{3.9}
\end{equation}
In equation (11), $U_{R}=[u_{1},u_{2},....u_{M}]$ can be shown as the eigenvector matrix and $\Lambda_{R}$ represents the eigenvalue matrix. The eigenvectors and eigenvalues will meet the following requirement:
\begin{equation}
Ru_{i}=\lambda_{i}u_{i}, i=1,2,....,M \label{3.10}
\end{equation}
Since the source signal and the noise are independent of each other, the spatial correlation matrix $R$ can be decomposed into two parts, which are the source signal and the noise. Thus, the spatial correlation matrix $R$ is eigendecomposed to obtain equation (13).
\begin{equation}
R=U_{R}\Lambda_{R}U_{R}^H=U_{s}\Lambda_{s}U_{s}^H+U_{N}\Lambda_{N}U_{N}^H
\label{3.11}
\end{equation}
$U_{s}$ is the subspace consisting of the larger eigenvectors among all eigenvalues of $R$, called the source signal subspace. $U_{N}$ is the subspace consisting of the smaller eigenvectors among all eigenvalues of $R$, named the noise subspace. $U_{s}$ is a matrix of $M \times K$ while $U_{N}$ is a matrix of $M\times(M-K)$. The eigenvector matrix will be utilized to build the spatial spectral function, which can be shown in equation (14).
\begin{equation}
P(\theta)=\frac{1}{\alpha^H(\theta)U_{N}U_{N}^H\alpha(\theta)}
\label{3.12}
\end{equation}
In equation (14), $\alpha$ represents the direction vector in the source signal subspace. The K largest peaks in the equation (14) correspond to the AoAs of the signals from K sources.

\subsection{ Hybrid Positioning Method based on RSSI and AoA}
To decrease the error and attain more precise receiver location information, a new hybrid positioning method utilizing RSSI values and AoA values is introduced into indoor positioning research. The measured AoA values will be utilized in the calculation of the coordinates of the receiver point.

The RSSI values from the three WiFi routers will be measured at the receiver point. The measured RSSI values are brought into the log-distance path loss model to calculate the distances from the WiFi routers to the receiver point. The distances from the three WiFi routers to the receiver point are measured as $d_{1}$, $d_{2}$, and $d_{3}$ respectively. Meanwhile, the MUSIC algorithm will be used to measure the AoA values of the signals from the three WiFi routers to the receiver point. The three  AoA values are measured as $\theta_{1}$, $\theta_{2}$ and $\theta_{3}$ respectively. The locations of the three WiFi routers are fixed, which can be demonstrated as $(x_{1},y_{1})$, $(x_{2},y_{2})$ and $(x_{3},y_{3})$ respectively. Finally, the measured RSSI and AoA values will be used to calculate the coordinates of the receiver point, which can be illustrated in the below equations.
\begin{equation}
x_{R1}=x_{1}+d_{1}\times\sin(\theta_{1})
\label{3.15}
\end{equation}
\begin{equation}
y_{R1}=y_{1}+d_{1}\times\cos(\theta_{1})
\label{3.16}
\end{equation}
\begin{equation}
x_{R2}=x_{2}+d_{2}\times\sin(\theta_{2})
\label{3.17}
\end{equation}
\begin{equation}
y_{R2}=y_{2}-d_{2}\times\cos(\theta_{2})
\label{3.18}
\end{equation}
\begin{equation}
x_{R3}=x_{3}-d_{3}\times\sin(\theta_{3})
\label{3.19}
\end{equation}
\begin{equation}
y_{R3}=y_{3}-d_{3}\times\cos(\theta_{3})
\label{3.20}
\end{equation}

To better determine the location of the receiver point, the average of $(x_{R1},y_{R1})$, $(x_{R2},y_{R2})$ and $(x_{R3},y_{R3})$ will be utilized to represent the coordinates of the receiving point. Theoretically, $(x_{R1},y_{R1})$, $(x_{R2},y_{R2})$ and $(x_{R3},y_{R3})$ are equal. However, noise and interference in the real indoor environments will have influence on the performance of the MUSIC algorithm, thus affecting the AoA values. The Fig.3 shows the outcome of the hybrid positioning method.

\begin{figure}[H] 
\centering 
\includegraphics[width=0.3\textwidth]{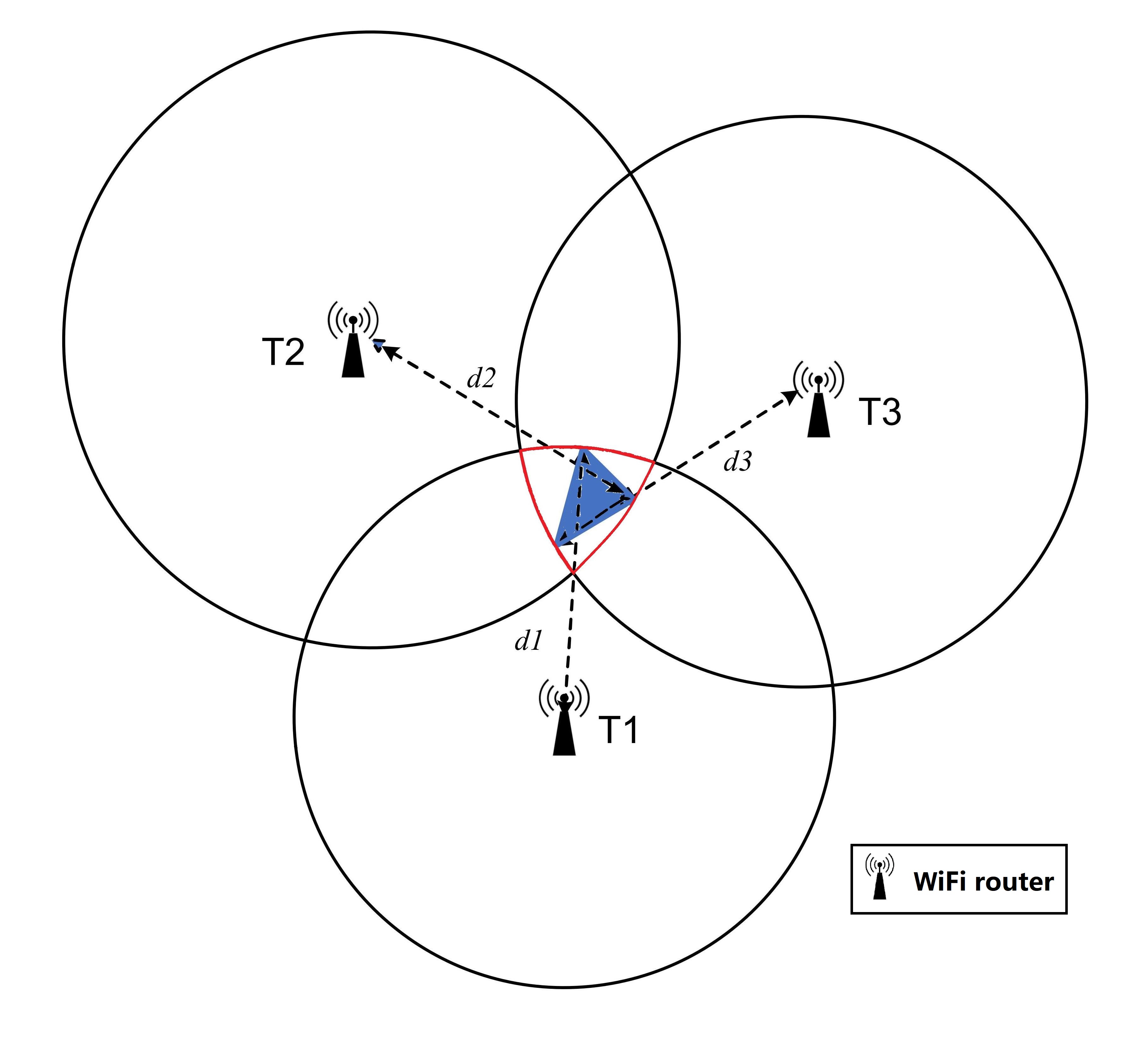} 
\caption{The Result of the Hybrid Positioning Method}
\label{Figure_3.5} 
\end{figure}

The location of the receiving point will be in the blue triangle. Compared with the RSSI-based Trilateration method, the hybrid positioning method reduces the error because the AoA values can provide more location information than the RSSI values.

\subsection{Deep Learning Methods}

Deep Learning algorithms have demonstrated good performance in regression, prediction, and classification of data. Neural networks reduce the NLoS problem and the impact of multipath effect by adjusting the weights of their internal neurons. In addition, neural networks can learn and extract features from the data in order to complete the computation of the data \cite{cheng2018localized}. Therefore, two models based on RSSI-based Trilateration method and hybrid positioning method will be built. The neural network can increase the accuracy of indoor positioning by training the model to obtain the optimal solution. The RSSI values will be utilized as the training data in the RSSI-based Trilateration model, while the RSSI values and AoA values will be utilized as the training data in the hybrid positioning model. The training targets
are the coordinates of the test points in the test environments.

There are three main steps by using the Deep Learning methods.

(1) Build the two models by using the reference points.

(2) Train the models by using Deep Learning algorithms.

(3) Test the performance and calculate errors.

In the first step, the fixed locations of the three WiFi routers will be used as the reference points in each test environment. Then train the model by using different Deep Learning algorithms and different results will be obtained. Three types of the Deep Learning algorithms will be used in this paper. They are BPNN, RBF and CNN. Finally, the test data sets will be utilized to assess the performance of algorithms. The mean absolute error (MAE) will be utilized to record the errors.

\section{Experiment}\label{sec:experiment}
\subsection{Experimental Environments}
To better evaluate the accuracy of the Deep Learning algorithms and explore the errors in different indoor environments, the experiments will be conducted in different indoor scenarios, including a big classroom, a corridor and a small classroom, as shown in Fig.4.
\begin{figure}[H] 
\centering 
\includegraphics[width=0.4\textwidth]{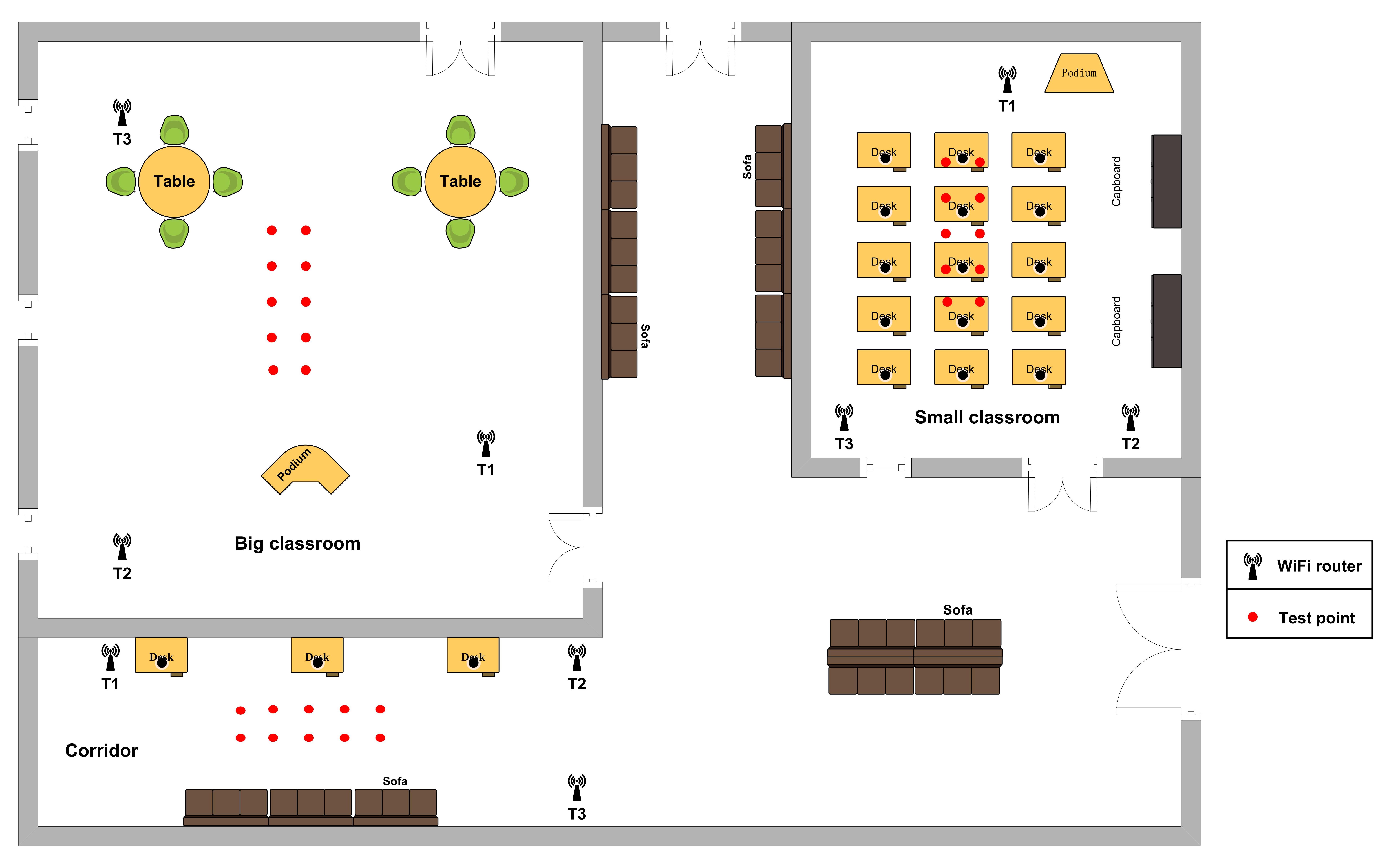} 
\caption{The Layout of the Test Environments}
\label{Figure_4.4} 
\end{figure}
The dimensions $(L\times W)$ of the big classroom are $13 m \times 13 m$. The dimensions $(L\times W)$ of the corridor are $12m \times 4m$, and the dimensions $(L\times W)$ of the small classroom are $9m\times 7m$. The three WiFi routers as the transmitters will be located at the corners of each test environment. The laptop and AD-FMCOMMS3-EBZ with the Zedboard will be used as the receiver. Ten test points are used as the locations of receiver. The coordinates of the WiFi routers and the test points will be demonstrated in Appendix A.
\subsection{Experimental Procedures}
The experimental procedures will be shown in the following steps.

(1) The path loss coefficient and log-normal fading variance will be calculated for each test environment. The values calculated in both directions for each test environment will be averaged, which would be illustrated in Table I and Table II.

\begin{table}[H]
\centering
\caption{The Path Loss Coefficients}
\includegraphics[width=0.45\textwidth,height=0.17\textwidth]{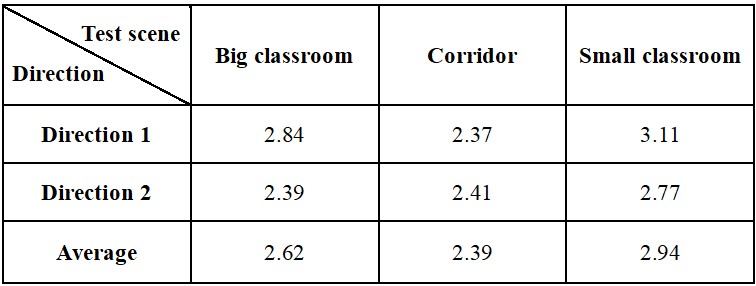}
\label{fig}
\end{table}

\begin{table}[H]
\centering
\caption{The Log-normal Fading Variances}
\includegraphics[width=0.45\textwidth,height=0.17\textwidth]{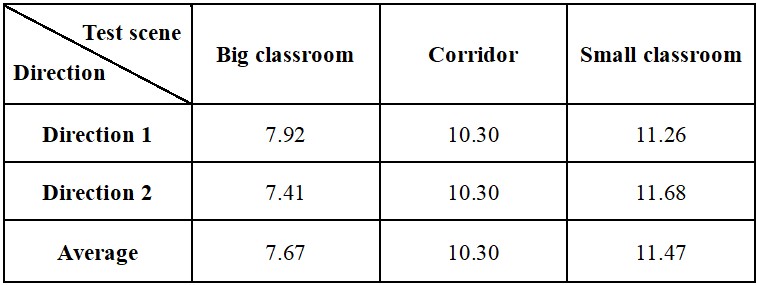}
\label{fig}
\end{table}
(2) The WiFi analyzer software on the laptop will be used to measure the RSSI values from the three WiFi routers. The theoretical RSSI values for test points will be calculated based on calculated distances, path loss coefficients and log-normal fading variances. A small percentage of the measured data is so different from the theoretical values that it is classified as outliers. Eliminating outliers not only improves the accuracy of Deep Learning algorithm models, but also reduces the negative impact of unreasonable data on experimental results. Therefore, when the absolute value of the difference between the theoretical RSSI value and the measured RSSI value is too large, the RSSI values need to be remeasured.

(3) The hardware together with the additional functions in MATLAB will capture the raw $I \& Q$ signal waveform. The raw $I \& Q$ signal waveform will be utilized as the input signal waveform of the MUSIC algorithm. The MUSIC algorithm will analyze the signal waveform and calculate the AoA values.

The theoretical values of the AoA can be calculated based on the coordinates of the WiFi routers and the receiver point. To decrease the negative influence of unreasonable experimental data on the experimental results, the AoA values need to be recalculated and remeasured when the absolute value of the difference between the measured value and the theoretical value is large.

(4) To better record the variation of RSSI values and AoA values for each test point, 500 groups of RSSI and AoA values will be collected for each test point as the training data set for the Deep Learning algorithms.

(5) The Deep Learning algorithms will be used to train the models and calculate the coordinates and errors. The whole procedure can be shown in Fig.5.
\begin{figure}[H] 
\centering 
\includegraphics[width=0.41\textwidth]{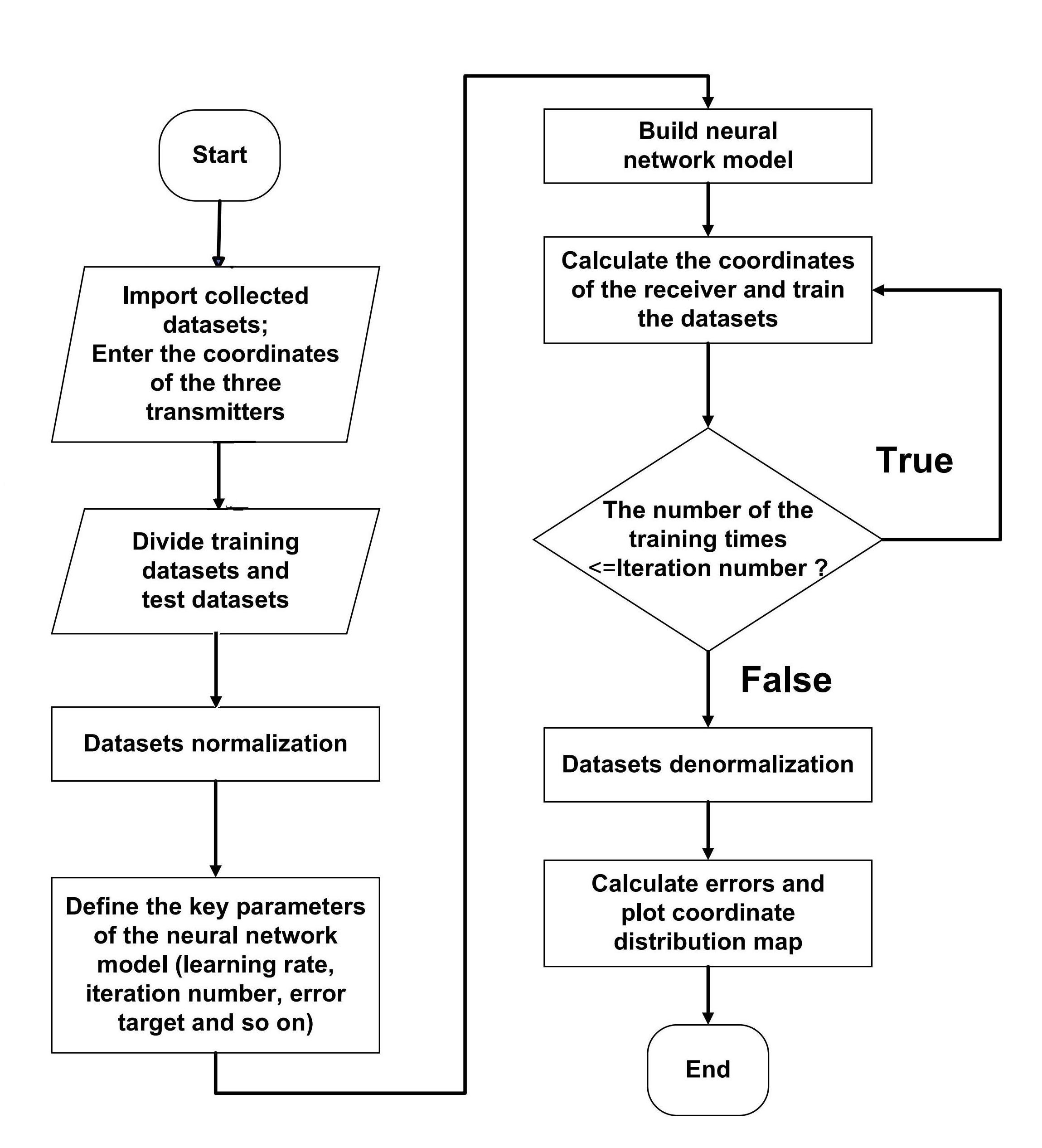} 
\caption{The Flow Chart of Neural Network}
\label{Figure_4.9} 
\end{figure}
First of all, the coordinates of the three WiFi routers will be used as reference points. Secondly, 80 $\%$ of the input data sets are divided into training data sets, while 20 $\%$ of the input data sets are defined as test data sets.  In order to reduce computational load and calculation time, the data sets will be normalized. In addition, after data normalization, data flattening will be employed in the CNN algorithm.  Thirdly, the parameters of the neural networks need to be defined. Then the neural networks will be constructed. The neural networks will reduce impacts of the NLoS problem and  multipath effect by adjusting the weights of neurons during the calculation. Multiple times of training help the neural networks to find the most suitable weights. The training will stop when the number of training times reach the defined number of iterations. The data sets need to be denormalized to return to the original ranges. Finally, the coordinates calculated by the Deep Learning algorithms will be compared with the theoretical values. The distance between the calculated coordinates and the theoretical coordinates will be defined as the error. The mean absolute errors (MAE) for all the test points in three test environments will be calculated.

\section{Results and Discussions}\label{sec:result and discussions}
\subsection{The Results of the RSSI-based Trilateration Model}
In this model, the WiFi analyzer will be used to measure the RSSI values from the three WiFi routers. The method in Section III will be used in the neural networks to calculate the coordinates of the test points. Meanwhile, the neural networks reduce the negative impacts of NLoS problem, multipath effect and other disturbance by adjusting the weights of neurons. The mean absolute error (MAE) will be used to assess the performance. The errors of three test environments can be shown in Fig.6 and Table III. To accurately describe the errors, the unit of error will be millimeters.

\begin{table}[H]
\centering
\caption{The MAEs of the RSSI-based Trilateration Model}
\includegraphics[width=0.45\textwidth,height=0.18\textwidth]{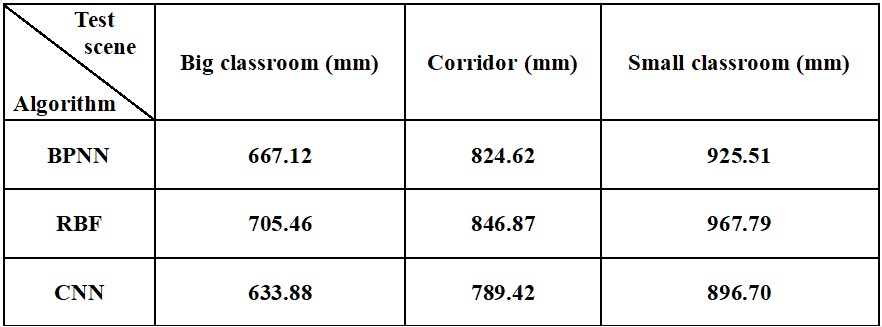}
\label{fig}
\end{table}

\begin{figure}[H] 
\centering 
\includegraphics[width=0.4\textwidth]{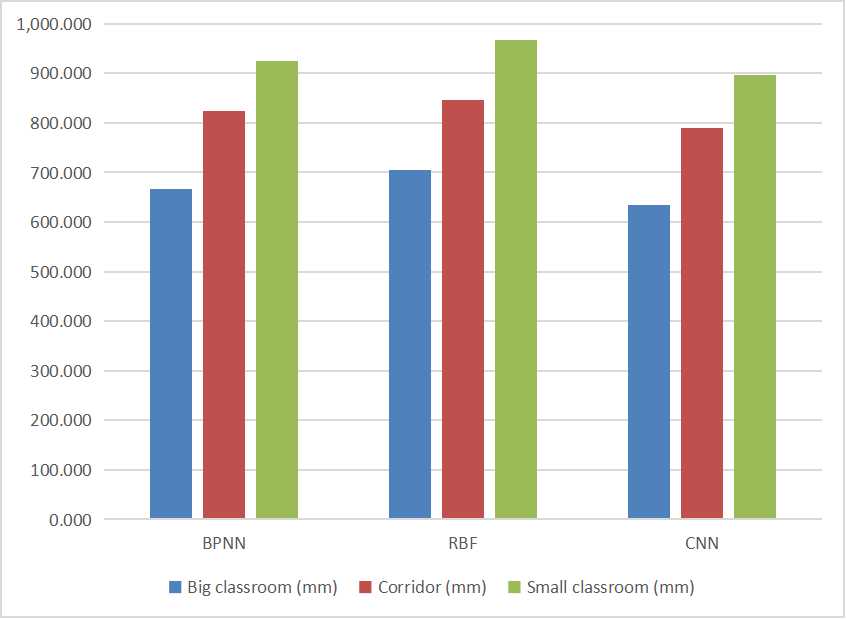} 
\caption{The MAEs of the RSSI-based Trilateration Model}
\label{Figure_4.9} 
\end{figure}
From Table III, the performance of CNN is the best compared to BPNN and RBF. Additionally, from Fig.6, the errors of the big classroom are the smallest, while the errors of the small classroom are the largest.  The big classroom has fewer obstacles. The dimensions of big classroom are large, so there are fewer signal reflections. However, there are more obstacles in the small classroom. The reflected signal travels a longer distance than the direct line signal to reach the receiving point due to the presence of obstacles. Therefore, NLoS problem leads to increased energy consumption during signal propagation, resulting in a lower measured RSSI value compared to the line-of-sight (LoS) propagation \cite{li2019performance,cai2020reconfigurable}. This can adversely affect the accuracy of positioning. 

In addition, WiFi analyzer is an open source software that obtains information about WiFi signals through a wireless network card. Wireless network cards are more focused on implementing wireless network connections rather than providing highly accurate signal measurements. As a result, WiFi analyzer may measure RSSI values with relatively low precision.

Compared with the traditional RSSI-based Trilateration approach in the paper \cite{rusli2016improved}, the precision of the RSSI-based Trilateration method combined with neural networks is improved. There are three reasons why the accuracy is increased. First of all, instead of relying on empirical values found on the websites, the path loss coefficients and log-normal fading variances calculated for different test environments  will be brought into equation (2) to calculate the distances from the WiFi routers to the test points. In addition, the RSSI data correction method will be used to clean up measured RSSI outliers to mitigate the negative effect of inaccurate experimental data on positioning accuracy. Most importantly, neural networks can adjust the weights of neurons to reduce the negative effects of noise and interference such as NLoS problem in the indoor environments. Neural networks also extract features from large amounts of data to identify and classify patterns in the data, accomplishing a large number of calculations.

\subsection{The Results of the Hybrid Positioning Model}
In this part, MUSIC algorithm will analyze the raw $I\&Q$ signal waveform captured by the AD-FMCOMMS3-EBZ with the Zedboard. The AoA values will be calculated. The RSSI values and the AoA values will be utilized as the input data of this model. The method in Section III combined with the Deep Learning algorithms will be used to calculate the coordinates of the test points. The neural networks decrease the negative influence of the noise and interference by changing the weights of the neurons among the output layers and the layers before the output layers. The errors of three test environments can be demonstrated in Fig.7 and Table IV.
\begin{table}[H]
\centering
\caption{The MAEs of the Hybrid Positioning Model}
\includegraphics[width=0.45\textwidth,height=0.18\textwidth]{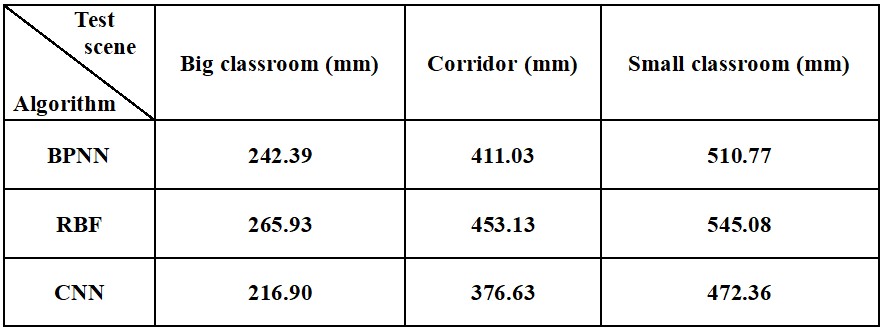}
\label{fig}
\end{table}

\begin{figure}[H] 
\centering 
\includegraphics[width=0.4\textwidth]{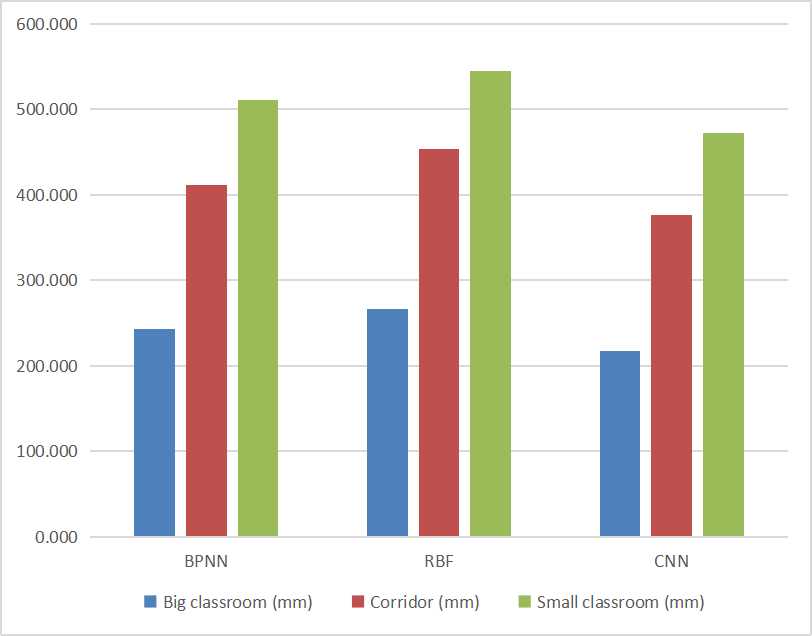} 
\caption{The MAEs of the Hybrid Positioning Model}
\label{Figure_4.9} 
\end{figure}
From the Table IV, the performance of CNN is the best one among the Deep Learning algorithms. Apart from it, the errors of the big classroom are the smallest, which are less than 300 mm. The errors of the small classroom are still big, which are approximately 500 mm. There are two main reasons why the errors of the small classroom are larger than those of the big classroom in this model. First of all, NLoS problem has a negative impact on AoA value measurements. In an indoor environment with many obstacles such as the small classroom, the signal strength will be attenuated due to the long propagation path and the presence of obstacles. The attenuated signal may result in increased noise in the measurement, which affects the accuracy of the AoA values. Secondly, in NLoS environments, the signal not only propagates through the direct path, but also reaches the test point through reflection and scattering. As a result, there will be a difference between the AoA of the signal after reflection and scattering, and the AoA of the signal traveling along the direct path. Therefore, the measured AoA values in a small classroom may be more different from the theoretical values, thus affecting the positioning accuracy \cite{ding2015performance,liu2018performance} .

Compared with the results of RSSI-based Trilateration model, the errors in three test environments are reduced, which indicates that the hybrid positioning method has a high accuracy. The hybrid positioning method can measure the AoA values in high accuracy. The AoA values provide more accurate positioning information, resulting in higher positioning accuracy. In addition, the MUSIC algorithm is a high-resolution algorithm that efficiently calculates the AoA values from the signal waveform. 

However, when three WiFi routers send signals at the same time, their signals are superimposed on each other at the receiving end. This may result in phase difference variations and interference effects that reduce the precision of the calculated AoA values \cite{zhai2017multi}. In addition, the movement of people and objects in indoor environments can also have a negative impact on the measurement of AoA values, thereby affecting the positioning accuracy.

\subsection{Comparison and Discussions}
By comparing the results in the previous two sections, it is clear that the errors of the hybrid positioning model are smaller than those of the RSSI-based Trilateration model, regardless of the test environments and the used Deep Learning algorithms. So the improvement between the two models in three different Deep Learning algorithms and three test environments will be calculated, which can be illustrated in Table V, Table VI and Table VII.
\begin{table}[H]
\centering
\caption{The Improvement in RBF}
\includegraphics[width=0.45\textwidth,height=0.18\textwidth]{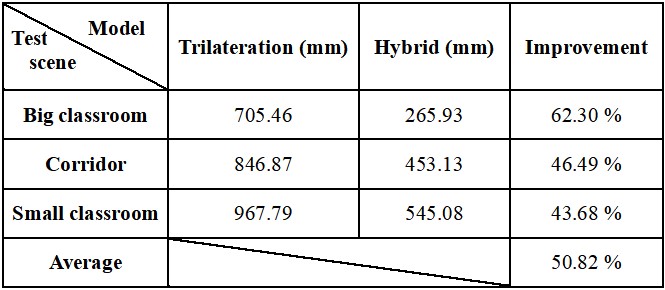}
\label{fig}
\end{table}

\begin{table}[H]
\centering
\caption{The Improvement in BPNN}
\includegraphics[width=0.45\textwidth,height=0.18\textwidth]{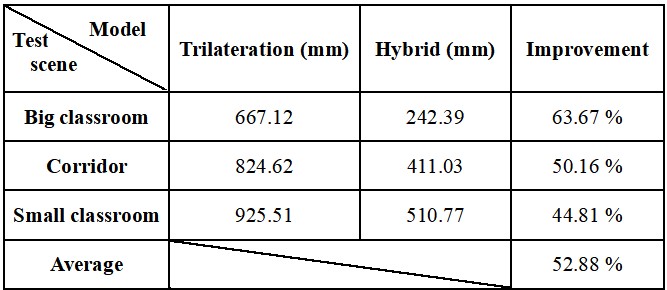}
\label{fig}
\end{table}

\begin{table}[H]
\centering
\caption{The Improvement in CNN}
\includegraphics[width=0.45\textwidth,height=0.18\textwidth]{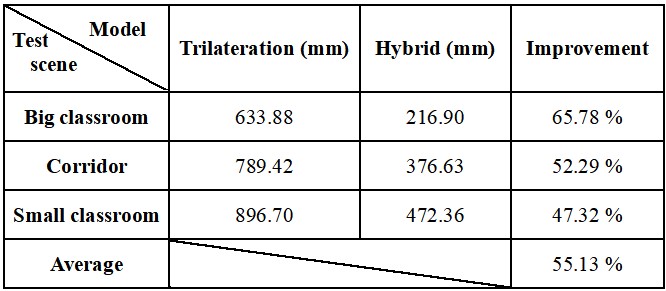}
\label{fig}
\end{table}

By analyzing from Table V to Table VII, the average improvement in the big classroom is greater than that in the small classroom.  The reason for this is that the NLoS problem has a less detrimental impact on the accuracy of the measured AoA values in a big classroom with fewer obstacles, compared to a small classroom with more obstacles. The accuracy of the measured AoA values in the big classroom is better than that in the small classroom, which can also have influence on the precision of indoor positioning. 

In addition, by comparing the improvement ability of the three Deep Learning algorithms, CNN has the best improvement ability, followed by BPNN, and RBF has the worst improvement ability.

Compared with RBF, CNN has the following advantages. First of all, CNN has the capabilities to automatically study and extract features from the input data by employing a combination of multi-layer convolution and pooling layers. In contrast, RBF networks usually require manual design and selection of appropriate Radial Basis functions for feature extraction \cite{xu2021performance}. Therefore, CNN is able to train on large-scale data by Back-Propagation algorithms, thus efficiently accomplishing a large number of computations. In addition, CNN has multiple fully connected layers but RBF has only one hidden layer. Therefore, CNN has more neurons than RBF. CNN can better reduce the influence of noise and interference on localization precision by adjusting the weights of neurons.

Although the improvement ability of BPNN is good, the improvement ability of CNN is better. First of all, CNN has the abilities to efficiently capture the local features of the input data through the use of convolutional operations and a parameter sharing mechanism. This makes CNN more advantageous when dealing with multidimensional data. In contrast, BPNN requires more parameters and computational resources to process these data . The dimension of the entered data for the RSSI-based Trilateration model is three, while the dimension of the entered data for the hybrid positioning model is six. So the errors of CNN are less than those of BPNN. In addition, CNN uses a multi-layer structure to study and pick up abstract characteristics of the entered data layer by layer. In contrast, BPNN has shallow, fully-connected networks that do not study the features layer by layer . The layered structure of CNN enables it to automatically learn hierarchical representations in the input data, thereby enhancing the model's capacity to comprehend the data \cite{lee2021comparing}.

Finally, BPNN has two advantages over RBF. First of all, BPNN can better capture complex nonlinear data relationships by utilizing nonlinear activation functions and multi-layer connections. In contrast, RBF has limited nonlinear fitting capability, which is more suitable for dealing with relatively simple data relationships \cite{ghose2010prediction}. The relationship between the RSSI values and the coordinates of test points is nonlinear. The relationship between the input items and output items in the hybrid positioning model is also nonlinear. The BPNN has better capabilities to process the nonlinear data, thus attaining a lower error. In addition, BPNN has multiple hidden layers, which can increase the learning capability of the network, while RBF has only one hidden layer. BPNN has more neurons than RBF, which can more effectively help BPNN to reduce the negative effects of NLoS problem and multipath effect by adjusting the weights of neurons \cite{ghose2010prediction}. Therefore, the learning ability of RBF is poor and it has a larger error.

\section{Conclusion}

In this paper, the WiFi indoor positioning has been investigated. Apart from the RSSI-based Trilateration method, a hybrid RSSI and AoA-based positioning method is innovated to improve the positioning accuracy. Additionally, to minimize the negative effects of noise and interference in indoor environments, neural networks will be used to train the RSSI-based Trilateration  model and the hybrid positioning model. The experimental results show that the errors of the hybrid positioning model are smaller compared to the RSSI-based Trilateration model. The errors obtained by utilizing CNN are the smallest among three Deep Learning algorithms. The errors in the big classroom are the smallest while the errors in the small classroom are the largest among the three test environments. The error of utilizing CNN to train the hybrid positioning model in the big classroom is the smallest, which is less than 250 mm. 

\section{Appendix.A}

\begin{table}[H]
\centering
\caption{The Coordinates of the Three WiFi Routers}
\includegraphics[width=0.47\textwidth,height=0.2\textwidth]{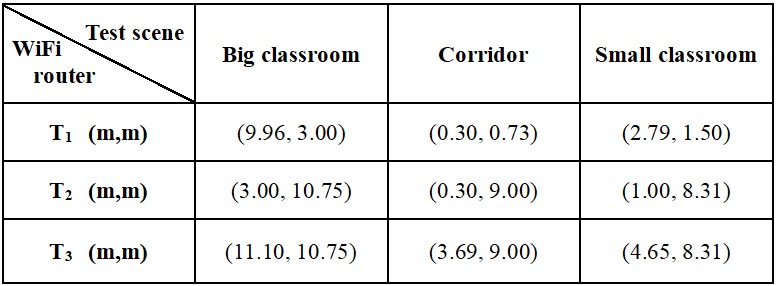}
\label{fig}
\end{table}

\begin{table}[H]
\centering
\caption{The Coordinates of the Test points in the Three Environments}
\includegraphics[width=0.45\textwidth,height=0.45\textwidth]{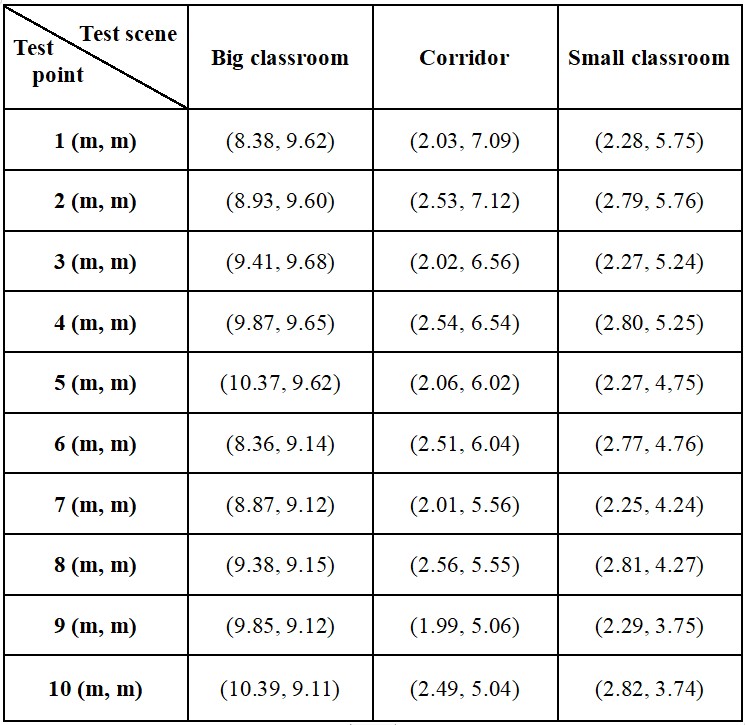}
\label{fig}
\end{table}


\bibliography{ref}


\end{document}